\documentclass[prl,aps,twocolumn,showpacs]{revtex4}
\usepackage{amsmath,amssymb}
\usepackage{graphicx}
\begin{document}

\title{Strong magnetoelastic effect in CeCo$_{1-x}$Fe$_{x}$Si as N\'eel order is suppressed}

\author{V. F. Correa}
\author{A. G. Villagr\'an Asiares}
\altaffiliation[Present address: ]{Nuklearmedizinische Klinik und Poliklinik Klinikum rechts der Isar der Technischen Universit\"at, M\"unchen, Germany}
\author{D. Betancourth}
\author{S. Encina}
\author{P. Pedrazzini}
\author{P. S. Cornaglia}
\author{D. J. Garc\'ia}
\author{J. G. Sereni}

\affiliation{Centro At\'omico Bariloche (CNEA) and Instituto Balseiro (U. N. Cuyo), 8400 Bariloche, R\'io Negro, Argentina}

\author{B. Maiorov}

\affiliation{National High Magnetic Field Laboratory, MS E536 Los Alamos National Laboratory, 
Los Alamos, New Mexico 87545, USA} 

\author{N. Caroca Canales}
\author{C. Geibel}

\affiliation{Max-Planck-Institut f\"ur Chemische Physik fester Stoffe, D-01187 Dresde, Germany}

\date{\today}


\begin{abstract}

A very strong magnetoelastic effect in the CeCo$_{1-x}$Fe$_{x}$Si alloys is reported. The strength of the magnetostrictive effect can be tuned upon changing $x$.
The moderate low-temperature linear magnetostriction observed at low Fe concentrations becomes very large ($\frac {\Delta L}{L} \left(16 T,2 K\right) =$ 3$\times$10$^{-3}$) around the critical concentration ($x_c \approx$ 0.23) at which the long-range antiferromagnetic order vanishes. Upon increasing doping through the non-magnetic region ($x > x_c$), the magnetostriction strength gradually weakens again.
Remarkably the low-temperature magnetostriction at the critical concentration shows a pronounced \textsl{S}-like shape (centered at $B_m \sim$ 6 T) resembling other well-known Ce-based metamagnetic systems like CeRu$_2$Si$_2$ and CeTiGe.
Unlike what is observed in these compounds, however, the field dependence of the magnetization shows only a minor upturn around $B_m$ vaguely resembling a metamagnetic behavior.
The subtle interplay between magnetic order and the Kondo screening seems to originate an enhanced valence susceptibility slightly changing the Ce ions valence, ultimately triggering the large magnetostriction observed around the critical concentration.

\end{abstract}

\maketitle

Magnetic order is usually discussed and described in terms of effective interactions between either localized and/or itinerant electrons \cite{Fazekas}. Even though the effective coupling can have a strong and non-trivial dependence on the distance or effective path between magnetic moments (as in the RKKY mechanism), the lattice effects are usually disregarded or just regarded as a second order property concomitant to the magnetic order, but not decisive to it \cite{Grazhdankina1969}.

The most notable exception to this general trend are probably the magnetic systems whose transitions to the ordered state are of first order type.  This fact turns out to be the signature that the magnetic transition occurs concurrently with an important lattice distortion. A strong volume dependence of the exchange couplings is the main responsible for the effect. Though the key role at such transitions is played by the magnetic interactions (not by the atomic lattice), yet it remains true that if there were no spin-lattice coupling the transition might occur at a different temperature or it could lose its first-order nature \cite{Grazhdankina1969}. In general, an important pressure dependence of the ordering temperature and a small bulk modulus are pre-requisites for the occurrence of first order magnetoelastic transitions \cite{Grazhdankina1969}.   

Another example of strong magnetoelastic effects occurs in the manganites. Even though the observed colossal magnetoresistance (CMR) around the magnetic order can be explained without invoking atomic lattice strain, it remains true that whenever there is CMR, there is also a giant magnetostrictive effect \cite{Ibarra1995}. In this case, the strong coupling is not only associated with a strain dependence of the super-exchange interaction \cite{Correa2012a} but it also involves atomic orbitals re-ordering associated with the Jahn-Teller effect \cite{Millis1995,Kimura1998}.

In this work we present another system where notable and strong magnetostructural effects are observed. The intermetallic CeCo$_{1-x}$Fe$_{x}$Si alloys show antiferromagnetic (AFM) ordering ($T_N$ = 8.8 K) in the stoichiometric limit ($x$=0) which weakens as the Fe content is increased \cite{Sereni2014}. The linear magnetostriction $\Delta L/L$, on the other hand, increases with $x$ approaching a maximum value of $\frac {\Delta L}{L} \left(16 T,2 K\right) =$ 3$\times$10$^{-3}$ at about the critical concentration $x_c \approx$ 0.23 where the magnetic order disappears. Beyond this doping level, the magnetostriction (MS) slowly but steadily decreases again as the system behaves as a heavy fermion. 
Also at the critical concentration, the MS displays a pronounced and hysteretic jump around $B_m \sim$ 6 T, very suggestive of a metamagnetic transition or crossover as it is observed in CeRu$_2$Si$_2$ (Ref. \cite{Puech1988}) and in the related CeTiGe (Ref. \cite{Deppe2012}). Unlike these compounds, the field dependence of the magnetization $M$ in CeCo$_{0.77}$Fe$_{0.23}$Si does not show the characteristic \textsl{S}-like shape of a metamagnetic transition, only a kink around $B_m$ above which the slope of $M(B)$ is almost doubled. Given the subtle interplay between the magnetic order and the Kondo screening, the sharp and large increase of the magnetostriction appears to be associated with the onset of a valence instability around $x_c$ which gives rise to a little change of the Ce effective valence across $B_m$. This valence change may also explain the large negative thermal expansion coefficient observed at high field. The interpretation is supported by the evolution of the unit-cell volume with the Fe content which shows an important Ce volume decrease above $x \sim x_c$ \cite{Sereni2014}, in line with previous magnetization \cite{Welter1992} and X-ray absorption spectroscopy \cite{Isnard1999} measurements where an important Ce valence change in CeFeSi relative to CeCoSi was reported.

High-quality single-phase polycrystalline samples of CeCo$_{1-x}$Fe$_{x}$Si used in this study were prepared by arc melting stoichiometric amounts of the pure elements followed by an annealing procedure as described previously \cite{Sereni2014}. 
A high-resolution capacitive dilatometer was used in the dilation experiments, while the magnetization measurements were carried out both in a SQUID magnetometer (up to 5 Tesla) and a VSM magnetometer (up to 14 Tesla). All dilation experiments under magnetic field where carried out in the longitudinal configuration, i.e. with the magnetic field $B$ parallel to the sample dimension $L$ being measured.
A standard heat-pulse technique was used in the specific heat experiments.

\begin{figure}[t]
\includegraphics[width=\columnwidth]{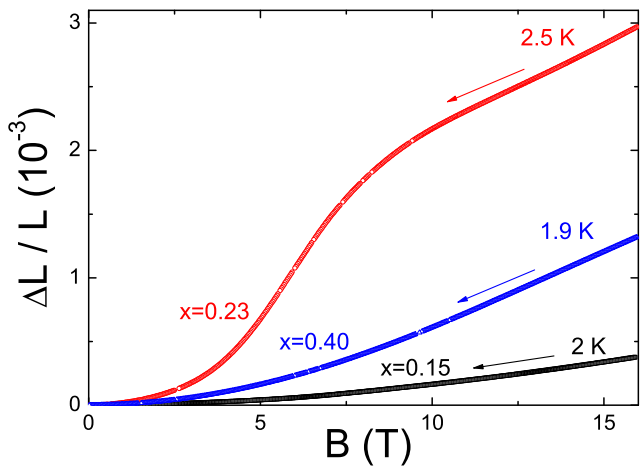}
\caption[]{(color online) Field dependence of the linear magnetostriction at $T \approx$ 2 K for different Fe concentrations. Arrows indicate the direction of the field sweeps.}
\label{fig1}
\end{figure}

Figure \ref{fig1} summarizes the main findings of this work. The low-temperature linear forced-magnetostriction (i.e., MS induced by the external field) is shown for three different Fe contents. A very large $\Delta L/L$ is seen at $x =$ 0.23 reaching a value as high as 3$\times$10$^{-3}$ at 16 Tesla. The effect is significantly reduced by a factor larger than 2 at $x =$ 0.4 and by one order of magnitude at $x =$ 0.15.
As reported in Ref. [\onlinecite{Sereni2014}], $x =$ 0.23 is approximately the critical concentration $x_c$ at which the antiferromagnetic order vanishes, while $x =$ 0.15 is placed well inside the magnetic region ($T_N \left[x=0.15\right]$ = 6.7 K) and $x =$ 0.4 is non-magnetic. In this sense, these $x$ values are representative of the different magnetic ground states observed.

\begin{figure}[t]
\includegraphics[width=\columnwidth]{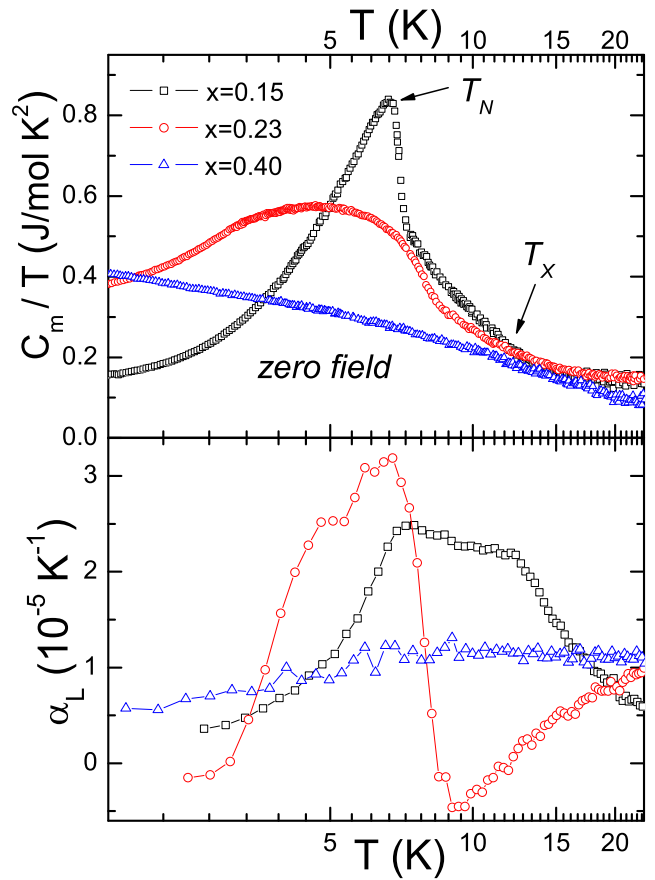}
\caption[]{(color online) Upper panel: magnetic contribution to the specific heat divided by temperature at selected Fe doping levels $x$ (adapted from Ref. [\onlinecite{Sereni2014}]). The lattice vibrations contribution has been substracted from the isotypic La compounds. Lower panel: linear thermal-expansion coefficient for the same Fe concentrations.}
\label{fig2}
\end{figure}

The upper panel of Fig. \ref{fig2} shows the low-temperature magnetic contribution to the specific heat ($C_m / T$) for the aforementioned concentrations. As we have shown in previous works \cite{Sereni2014,Correa2016}, the magnetic transition at $T_N$ is preceded by a large tail whose onset is at $T_X$. With increasing $x$, this tail grows up continuously, evolving into a large bump anomaly as the magnetic order collapses around $x_c$. At higher $x$, even the bump anomaly disappears as it can be seen in Fig. \ref{fig2}. 

The lower panel of Fig. \ref{fig2} shows the corresponding linear thermal-expansion coefficient $\alpha_L$ (=$\alpha_V$/3, the volume thermal-expansion coefficient, given the non-textured nature of the polycrystalline samples) for the selected concentrations. As we pointed out previously \cite{Correa2016}, the tail in the specific heat at $x=$0.15 exhibits a remarkably large coupling to the lattice as shown by the double peak structure in $\alpha_L$, which is consistent with the presence of a structural transition preceding the magnetic transition according to a mean field model \cite{William2018}.
This large spin-lattice coupling is now further confirmed by the thermal-expansion at $x=$0.23. While the magnetic order is almost suppressed, $\alpha_L$ shows a broad and large bump (following a pronounced minimum) at basically the same temperature where the bump in the specific heat is observed. Then, at $x=$0.4, $\alpha_L$ is largely reduced.

\begin{figure}[t]
\includegraphics[width=\columnwidth]{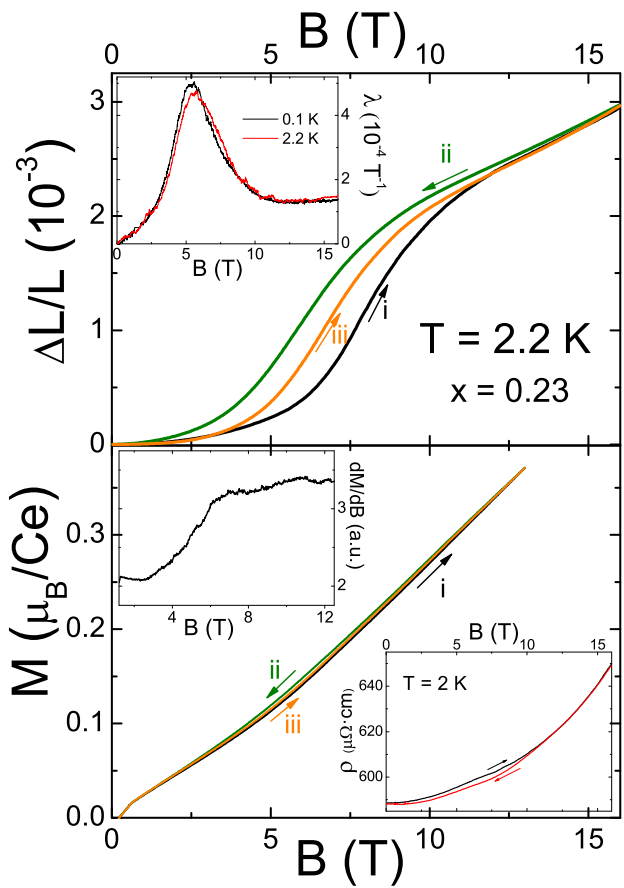}
\caption[]{(color online) Upper panel: linear forced-magnetostriction at $T =$ 2.2 K; inset: magnetostriction coefficient $\lambda = \frac{\partial \Delta L/L}{\partial B}$ at two different temperatures. Lower panel: field dependence of the magnetization at $T =$ 2.2 K; left inset: field derivative of the magnetization; right inset: magnetoresistivity at $T =$ 2 K. All data for $x = 0.23 \approx x_c$ . See text for details about curves i, ii, iii. Arrows indicate the direction of the field sweeps.}
\label{fig3}
\end{figure}

Two factors are at play in the suppression of the antiferromagnetic phase as the Fe concentration is increased. On the one hand, there is an increased hybridization of the magnetic moments on the Ce ions \textit{f} shell with the conduction band which leads to a screening of the magnetic moments through Kondo physics. On the other hand, since these 111 compounds can be described as a stacking of rare earth (Ce), transition metal (Co, Fe) and semimetal (Si) layers \cite{Sereni2014}, the substitution of a transition metal atom is expected to change the interlayer interaction between the magnetic moments in Ce ions, introducing random links (disorder) in the couplings between planes. A simple double-exchange argument indicates that the nearest neighbor  interlayer magnetic interaction will change sign because Co and Fe differ by a single electron in the \textit{d} shell. This change in the sign of the interactions is consistent with the magnetic behavior of GdCo$_{1-x}$Fe$_x$Si \cite{Nikitin1996,Wlodarczyk2015,VildasolaFuturo} and CeTi$_{1-x}$Sc$_x$Ge compounds \cite{Sereni2015}.

The very large forced-magnetostriction observed in Fig. \ref{fig1} at $x= 0.23$ together with the also considerable spontaneous-magnetostriction (i.e., zero field MS induced by the magnetic correlations) seen in Fig. \ref{fig2} at the same concentration points towards the magnetic origin of the bump anomaly, most likely associated with developing short-range correlations.
On the other hand, strong hysteresis occurs around $x_c$. This is shown in the upper panel of Fig. \ref{fig3} where three consecutive magnetostriction field-sweeps curves at $T \approx$2 K are displayed. Curve (i) stands for the first up-sweep after zero field cooling, while curves (ii) and (iii) are subsequent down-sweep and up-sweep, respectively.  
This hysteresis is also observed at $x=$0.15, though smaller\cite{Correa2016}, and it becomes negligible at $x=$0.4 (not shown here).    
It is intriguing, however, that the hysteresis in the magnetization is much smaller, as can be seen in the lower panel of Fig. \ref{fig3} where curve labeling follows that of magnetostriction.
On the other hand, no hysteresis in the magnetization is observed at $x=$0.15 nor at $x=$0.4.

\begin{figure}[t]
\includegraphics[width=\columnwidth]{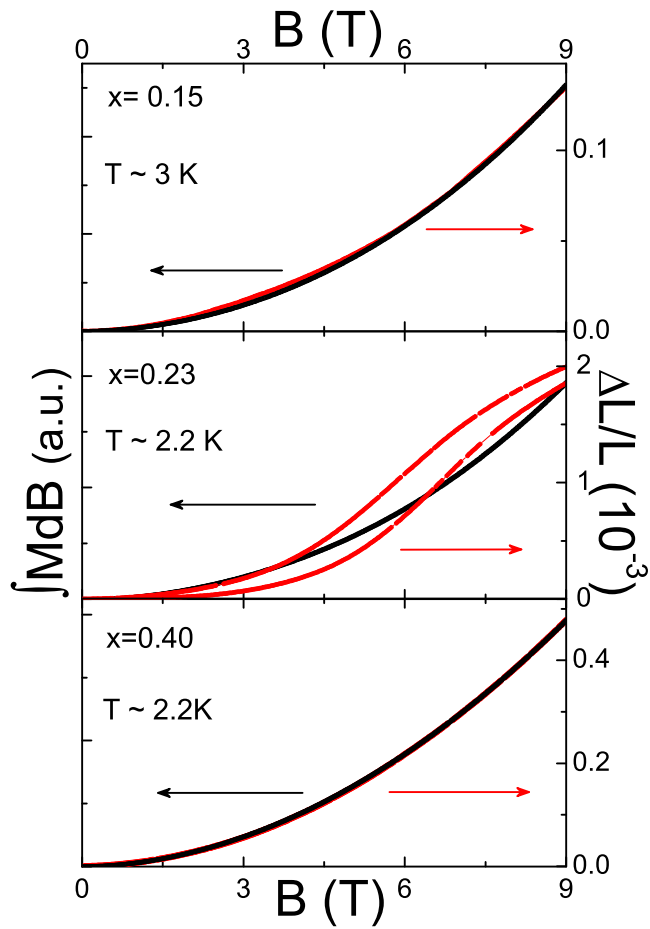}
\caption[]{(color online) Comparison between the integral $\int \! M \, dB$ (left vertical axis) and $\Delta L(B) / L$ (right vertical axis) at low temperature and for the three selected concentrations: $x =$ 0.15, 0.23 and 0.4 in the upper, middle and lower panel, respectively. See text for details.}
\label{fig4}
\end{figure}

The non-trivial coupling between lattice and spin degrees of freedom and the key role played by the lattice around $x_c$ is further revealed by the following analysis using thermodynamic Maxwell relationships. Considering an appropriate free energy $G(T,P,B)$ in its simplest form such that $dG(T,P,B) = -SdT + VdP - mdB$, the relationship $\left( \frac{\partial m}{\partial P} \right)_{T,B} = -\left( \frac{\partial V}{\partial B} \right)_{T,P}$ is implied, where $m$ is the magnetic moment. Integrating we get 

\begin{equation}
\int \frac{\partial M}{\partial P} dB = - \frac{\Delta V(B)}{V} 
\end{equation}
 
\noindent 
for a given temperature and pressure.
Figure \ref{fig4} is a comparative plot between the linear forced-magnetostriction $\Delta L(B) / L$ (we do not have the actual volume effect since we have not measured the transverse MS) and the integral $\int \! M \, dB$ for the three selected concentrations. Both magnitudes show a perfect correlation at $x =$ 0.4 (lower panel). This is something that we could have anticipated given the vanishing character of the magnetic correlations and the linear field dependence of the magnetization in the paramagnetic regime. It is nothing but the prediction of the magnetoelastic theory \cite{Wohlfarth1969}, $\Delta V(B) / V \propto M^2$, expressed in a different form.
Interestingly, a quite good correlation is also observed in the magnetic regime (see the upper panel for $x =$ 0.15) despite the fact we are integrating $M$, not its derivative $\partial M / \partial P$. 
The correspondence, however, definitely breaks down at $x =$ 0.23 as seen in the middle panel of Fig. \ref{fig4}. 
Even at higher temperature ($T \approx 10$ K, not shown here) where the magnetic fluctuations are weaker the correlation is far from good.

Another interesting aspect to remark is the shape of the magnetostriction curve at $x =$ 0.23 (upper panel of Fig. \ref{fig3}). It resembles that of a metamagnetic crossover as it is observed, for instance, in CeRu$_2$Si$_2$ (Ref. \cite{Puech1988}) or, more recently, in the relative compound CeTiGe, which shows a pronounced first-order metamagnetic transition \cite{Deppe2012} around $B_m \approx$ 12 T. 

A comparison with CeTiGe is particularly worthwhile. The longitudinal linear forced-magnetostriction displays many similarities with CeCo$_{0.77}$Fe$_{0.23}$Si: (i) the characteristic S-like shape of a metamagnetic transition; (ii) an important hysteresis around $B_m$; (iii) a similar and very large value $\Delta L(16 T) / L \sim$ 3$\times$10$^{-3}$ in both compounds; (iv) a step-like increase $\Delta L(B_m) / L \sim$ 2$\times$10$^{-3}$ at the transition; (v) temperature independence of $\Delta L/ L$ below a few Kelvin (see inset in the upper panel of Fig. \ref{fig3} of this work and Fig. 4 in Ref. \onlinecite{Deppe2012}).  

Nevertheless, there are important differences as well. The most important and astonishing is that the usual correlation between magnetization and large magnetostriction observed in metallic metamagnets is missing in CeCo$_{0.77}$Fe$_{0.23}$Si. The magnetization shows no clear indication of a metamagnetic effect as seen in Fig. \ref{fig3} (just a kink at $B_m \sim$ 6 T where the magnetostriction shows the pronounced increase). On the other hand, CeTiGe shows a sizeable magnetization jump close to 1 $\mu_B / Ce$ at $B_m$ which manifests itself even in the electrical resistivity as a pronounced decrease. In CeCo$_{0.77}$Fe$_{0.23}$Si, only a tiny bump is  observed in the magnetorresistance around $B_m$, as seen in the lower inset of Fig. \ref{fig3}. 
A very good correlation is also observed in CeRu$_2$Si$_2$ where thermodynamic analysis similar to what we have made in Fig. \ref{fig4} gives an excellent correspondence between magnetization and magnetostriction, even around $B_m$ \cite{Puech1988}.  

\begin{figure}[t]
\includegraphics[width=\columnwidth]{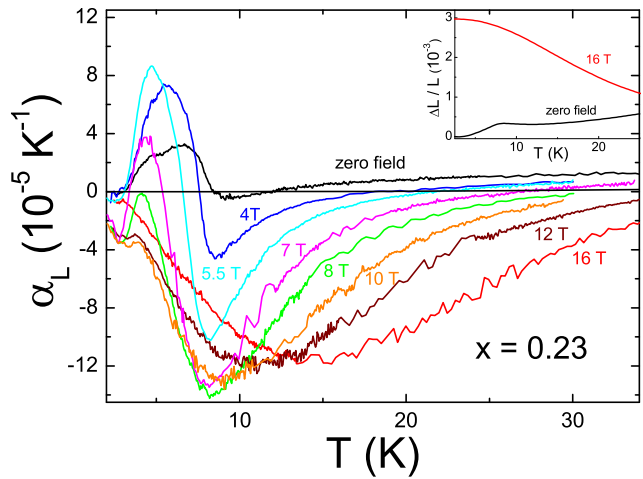}
\caption[]{(color online) Linear thermal-expansion coefficient for x = 0.23 at different applied magnetic fields.}
\label{fig5}
\end{figure}

Both CeTiGe and CeRu$_2$Si$_2$ show another common observation in metamagnetic systems: a sign change of the thermal expansion across the metamagnetic characteristic field \cite{Deppe2012,Lacerda1989}. On the other hand, this effect is not clear in CeCo$_{0.77}$Fe$_{0.23}$Si. Figure \ref{fig5}  displays the temperature dependence of the linear thermal expansion coefficient $\alpha_L$ at different applied magnetic fields. 
Though $\alpha_L$ evolves from positive to negative as the field is raised, it is not clear that the sign change occurs at $B_m$. 
The effect could be masked by the large hysteresis observed which indeed prevents a definition of the exact value of $B_m$. 
Besides this, and unlike CeTiGe and CeRu$_2$Si$_2$, the residual magnetic interactions are still quite strong which manifest themselves in the prominent bump-structures in $\alpha_L$, mainly at low fields.
Another aspect to stress from Fig. \ref{fig5} is that as long as the field is increased above $B_m$ and the bump structure washes out, the overall shape of $\alpha_L$ is very reminiscent of a Schottky anomaly due to isolated magnetic moments. In fact, the temperature $T_m$ at which $\alpha_L$ is minimum has a linear dependence with the magnetic field, above 8 T.  

\begin{figure}[t]
\includegraphics[width=\columnwidth]{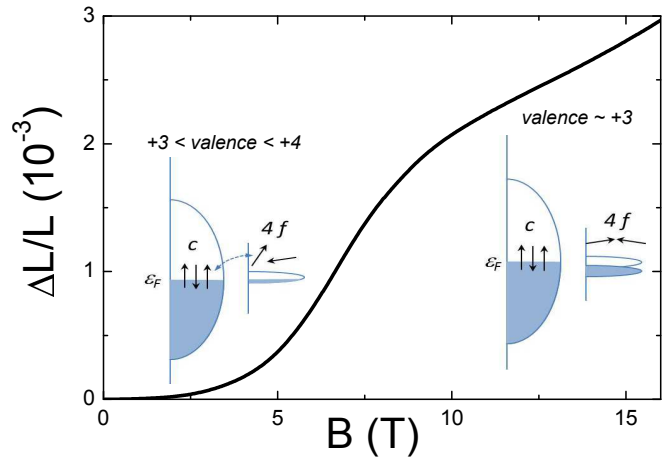}
\caption[]{(color online) Schematic picture of the magnetostriction at the critical concentration. The external magnetic field induces a transition from a Ce mixed-valence state to a Ce$^{3+}$ predominant state.}
\label{fig6}
\end{figure}

What could the origin of the large \textsl{S}-shaped magnetostriction at $x_c$?
The concentration $x_c$ is at the onset of the Ce volume Kondo collapse, as reported in Ref. [\onlinecite{Sereni2014}]. This is in agreement with previous magnetization \cite{Welter1992} and X-ray absorption spectroscopy \cite{Isnard1999} measurements which clearly show that Ce$^{3+}$ is the electronic state in CeCoSi, while CeFeSi is in an intermediate valence state. Hence, the large magnetostriction around $x_c$ seems to be the consequence of an incipient valence instability whose onset is around $x_c$. At lower Fe concentrations, the 4$f$ level is below the Fermi level $\epsilon_F$ and the Ce$^{3+}$ moments are ordered. As $x$ increases, the 4$f$ level approaches the Fermi level triggering the hybridization with the conduction band, gradually entering into a mixed-valence state (mixture of Ce$^{3+}$ and Ce$^{4+}$), thus suppressing the magnetic order while evolving into a Kondo state cleary observed at higher $x$. 
Around $x_c$ the very narrow 4$f$ band and $\epsilon_F$ are close enough as to start having considerable charge fluctuations. The system is expected to be particularly susceptible to an external magnetic field. Hence, the incipient non-orderder and partially non-localized mixed-valence state can be turned energetically unfavorable under a moderate magnetic field and an ordered (may be a canted AFM to take advantage of the Zeeman energy), localized 3+ valence state can be reinstated. The large volume change difference between the Ce$^{4+}$ and Ce$^{3+}$ configurations explains then the large lattice change at $B_m$. The situation is depicted schematically in Fig. \ref{fig6}. Indeed, one can estimate a change $\Delta L / L \sim 10^{-3} $ from the evolution of the lattice parameters with $x$ according to Ref. \onlinecite{Sereni2014} supposing that the lattice volume of CeCoSi is recovered upon the application of a magnetic field (after subtracting the intrinsic expansion considering the analog series LaCo$_{1-x}$Fe$_{x}$Si). This should be compared with the jump seen at $B_m$, which is of the same order.
Concomitantly, and though $M$ shows no abrupt change at $B_m$, its field derivative (i.e., susceptibility) does show a jump at $B_m$, as observed in the left inset of the lower panel of Fig. \ref{fig3}.

This valence change picture offers also a possible explanation for the Schottky-like behavior of the thermal-expansion coefficient (Fig. \ref{fig5}). Around $x_c$, the width of the narrow $f$-bands should be order $\sim B_m$. For $B \geq B_m$ one may expect a full split between the spin-up and spin-down bands, with this last one being nearly depopulated. 
In this intuitive scenario, these two narrow bands can be seen as a two-level system which gives rise to the Schottky anomaly in $\alpha_L$. 
The negative thermal expansion is also consistent with this view: the large MS is suppressed as long as the temperature is increased (see inset of Fig. \ref{fig5}) because the spin-down band is populated and the two-level picture is washed out. Because the hybridization with the conduction band should raise, the Ce ions should slightly lose their 3+ character.  

Finally, it is noteworthy to mention that a partial suppression of the hybridization between conduction and 4$f$ electrons (i.e., suppression of the Kondo effect) by an applied magnetic field has already been observed in other Ce-based compounds, like Ce$_{0.8}$La$_{0.1}$Th$_{0.1}$ \cite{Drymiotis2005}.

\section{Conclusions}

A strong magnetoelastic effect is reported in the CeCo$_{1-x}$Fe$_{x}$Si alloys. The forced magnetostriction $\Delta L(B) / L$ is shown to change by an order of magnitude in response to slight changes of the Fe content $x$ showing a maximum around the critical concentration $x_c$ where the N\'eel order is suppressed. At this critical concentration, the magnetostriction shows a $S$-like shape very reminiscent of a metamagnetic behavior. 

Given the subtle interplay between the magnetic order and the Kondo screening, the large magnetostriction appears to be associated with the onset of a valence instability around $x_c$ with the magnetic field reversing the mixed-valence state towards a localized 3+ state thus giving rise to a large volume change. This interpretation is supported by the evolution of the unit-cell volume with $x$ which confirms an important Ce volume reduction \cite{Sereni2014}, in agreement with magnetization \cite{Welter1992} and X-ray absorption spectroscopy \cite{Isnard1999} measurements that show a considerable Ce valence change between CeCoSi and CeFeSi. The Schottky-like shape shown by the large and negative thermal-expansion coefficient at high field is also consistent with this valence change scenario.

At $x_c$, the magnetostriction also shows an important hysteresis which is basically absent in the magnetization. This would imply a strong pinning mechanism acting on the atomic lattice but not on the magnetic moments.

The magnetostrictive effect maximum around $x_c$ is something we could have anticipated given the close competition between different energy scales. Indeed, it has been predicted and demonstrated in CeRu$_2$Si$_2$ (the paradigmatic example of a Kondo system in the very border of a magnetic instability) upon small substitutions of Ce by La \cite{Lacerda1988} or Ru by Rh \cite{Takeuchi1996}. However, the effect is not as evident as in CeCo$_{1-x}$Fe$_{x}$Si.

\section{Acknowledgments}

The authors gratefully acknowledge helpful discussions with C. D. Batista.
V.F.C, P.P, P.S.C, D.J.G and J.G.S are members of CONICET, Argentina. Work performed in Bariloche is partially supported by ANPCyT PICT2010-1060 Bicentenario and PICT2016-0204, PIP 112-2013-0100576, SeCTyP-UNCuyo 06/C513 and 06/C520.
B.M. was supported by the U.S. DOE, Office of Science, BES, Materials Sciences and Engineering Division. Work performed at NHMFL Pulsed Field facility at LANL is supported by the National Science Foundation through NSF/DMR-1157490/1644779 and the State of Florida.


\vspace{15pt}

\begin{thebibliography}{99}

\bibitem{Fazekas} P. Fazekas, in \textit{Lecture Notes on Electron Correlation and Magnetism} (Singapore: World Scientific, 2003).

\bibitem{Grazhdankina1969}  N. P. Grazhdankina, Sov. Phys. Usp. \textbf{11}, 727 (1969).

\bibitem{Ibarra1995}  M. R. Ibarra, P. A. Algarabel, C. Marquina, J. Blasco, and J. Garc\'ia, Phys. Rev. Lett. \textbf{75}, 3541 (1995).

\bibitem{Correa2012a}  V. F. Correa, N. Haberkorn, G. Nieva, D. J. Garc\'ia, and B. Alascio, Europhys. Lett. \textbf{98}, 37003 (2012).

\bibitem{Millis1995}  A. J. Millis, P. B. Littlewood, and B. I. Shraiman, Phys. Rev. Lett. \textbf{74}, 5144 (1995).

\bibitem{Kimura1998}  T. Kimura, Y. Tomioka, A. Asamitsu, and Y. Tokura, Phys. Rev. Lett. \textbf{81}, 5920 (1998). 

\bibitem{Sereni2014}  J. G. Sereni, M. G\'omez Berisso, D. Betancourth, V. F. Correa, N. Caroca Canales, and C. Geibel, Phys. Rev. B \textbf{89}, 035107 (2014).

\bibitem{Puech1988}  L. Puech, J. M. Mignot, P. Lejay, P. Haen, J. Flouquet, and J. Voiron, J. Low Temp. Phys. \textbf{70}, 237 (1988).

\bibitem{Deppe2012}  M. Deppe, S. Lausberg, F. Weickert, M. Brando, Y. Skourski, N. Caroca-Canales, C. Geibel, and F. Steglich, Phys. Rev. B 85, 060401(R) (2012).

\bibitem{Welter1992}  R. Welter, G. Venturini, and B. Malaman, J. Alloys Compd. \textbf{189}, 49 (1992).

\bibitem{Isnard1999}  O. Isnard, S. Miraglia, R. Welter, and B. Malaman, J. Synchrotron Rad. \textbf{6}, 701 (1999). 

\bibitem{Correa2016}  V. F. Correa, D. Betancourth, J. G. Sereni, N. Caroca Canales, and C. Geibel, J. Phys.: Condens. Matter \textbf{28}, 346003 (2016).

\bibitem{William2018}  W. G. Carreras Oropesa, V. F. Correa, J. G. Sereni, D. J. Garc\'ia, and P. S. Cornaglia, J. Phys.: Condens. Matter \textbf{30}, 295803 (2018).

\bibitem{Nikitin1996} S. A. Nikitin, T. I. Ivanova, I. G. Makhro, and Yu. A. Tskhadadze, J. Magn. Magn. Mater. \textbf{157-158}, 387 (1996).

\bibitem{Wlodarczyk2015}  P. Wlodarczyk, L. Hawelek, P. Zackiewicz, T. Rebeda Roy, A. Chrobak, M. Kaminska, A. Kolano-Burian, and J. Szade, Mater. Chem. Phys. \textbf{162}, 273 (2015).

\bibitem{VildasolaFuturo} V. Vildosola \textit{et al.}, to be published.

\bibitem{Sereni2015}  J. G. Sereni, P. Pedrazzini, M. G\'omez Berisso, A. Chacoma, S. Encina, T. Gruner, N. Caroca-Canales, and C. Geibel, Phys. Rev. B \textbf{91}, 174408 (2015).

\bibitem{Wohlfarth1969}  E. P. Wohlfarth, J. Phys. C \textbf{2}, 68 (1969).

\bibitem{Lacerda1989}  A. Lacerda, A. de Visser, L. Puech, P. Lejay, P. Haen, J. Flouquet, J. Voiron, and F. J. Okhawa, Phys. Rev. B 40, 11429(R) (1989).

\bibitem{Lacerda1988}  A. Lacerda, A. de Visser, L. Puech, P. Lejay, and P. Haen, J. Magn. Magn. Mater. \textbf{76-77}, 138 (1988).

\bibitem{Takeuchi1996}  T. Takeuchi and Y. Mikayo, J. Phys. Soc. Jpn. \textbf{65}, 3242 (1996).

\bibitem{Drymiotis2005}  F. Drymiotis, J. Singleton, N. Harrison, J. C. Lashley, A. Bangura, C. H. Mielke, L. Balicas, Z. Fisk, A. Migliori, and J. L. Smith, J. Phys.: Condens. Matter \textbf{17}, L77 (2005).

\end{thebibliography}
\end{document}